\documentclass[11pt]{article}

\newcommand{\rom}[1]{\mathrm{#1}}

\usepackage{amsmath,amssymb,graphicx}
\usepackage{hyperref}

\usepackage{multicol,color,longtable}

\definecolor{darkred}{rgb}{0.65,0.15,0}
\hypersetup{pdfborder={0 0 0},colorlinks=true,urlcolor=darkred,citecolor=blue,linkcolor=darkred,linktocpage=true}

\definecolor{AV}{rgb}{0.65,0.0,0}
\definecolor{AK}{rgb}{0,0,1}
\definecolor{DK}{rgb}{0.6,0.4,0}

\newcommand{\sKerr}{\sigma_{\mathrm{Kerr}}}

\newcommand{\nn}{\nonumber}

\newcommand{\reals}{\mathbb{R}}

\newcommand{\cV}{\mathcal{V}}

\newcommand{\cM}{\mathcal{M}}

\newcommand{\id}{1\!\!1}

\newcommand\no{\nonumber}

\newcommand \RR{{\mathbb{R}}}

\newcommand\be{\begin{equation}}
\newcommand\ee{\end{equation}}

\newcommand\bea{\begin{eqnarray}}
\newcommand\eea{\end{eqnarray}}

\def\no{\nonumber}
\newcommand{\beq}{\begin{eqnarray}}
\newcommand{\eeq}{\end{eqnarray}}

\def \RR{{\mathbb{R}}}

\def\a{\alpha}
\def\b{\beta}
\def\g{\gamma}
\def\h{\eta}
\def\G{\Gamma}
\def\s{\sigma}

\makeatletter

\@addtoreset{equation}{section}
\makeatother

\begin{document}

\thispagestyle{empty}

{\flushright {AEI-2013-265}\\[15mm]}

\begin{center}
{\LARGE \bf An Inverse Scattering Formalism \\[3mm] for STU Supergravity}\\[10mm]

\vspace{8mm}
\normalsize
{\large  Despoina Katsimpouri${}^{1}$, Axel Kleinschmidt${}^{1,2}$\\[2mm] and Amitabh Virmani${}^3$}

\vspace{10mm}
${}^1${\it Max-Planck-Institut f\"{u}r Gravitationsphysik (Albert-Einstein-Institut)\\
Am M\"{u}hlenberg 1, DE-14476 Potsdam, Germany}
\vskip 1 em
${}^2${\it International Solvay Institutes\\
ULB-Campus Plaine CP231, BE-1050 Brussels, Belgium}
\vskip 1 em
${}^3${\it Institute of Physics\\
Sachivalaya Marg, Bhubaneshwar, Odisha, India 751005}

\vspace{15mm}

\hrule

\vspace{10mm}

\begin{tabular}{p{12cm}}
{\small

STU supergravity becomes an integrable system for solutions that effectively only depend on two variables.
This class of solutions includes the Kerr solution and its charged generalizations that have been studied in the literature.
We here present an inverse scattering method that allows to systematically construct solutions of this integrable system.
The method is similar to the one of Belinski and Zakharov for pure gravity but uses a different linear system due to Breitenlohner and Maison and here requires some technical modifications.
We illustrate this method by constructing a four-charge rotating solution from flat space. A generalization to other set-ups is also discussed.

}
\end{tabular}
\vspace{5mm}
\hrule
\end{center}

\newpage
\setcounter{page}{1}

\tableofcontents

\section{Introduction}

The method of inverse scattering, pioneered in gravity by Belinski and Zakharov~\cite{BZ1, BZ2, BV}, has been applied very successfully to pure gravity in $D=4$ and $D=5$ space-time dimensions
(see also the reviews~\cite{Emparan, Iguchi, Rocha:2013qya}). The method rests on identifying a linear set of equations with a spectral parameter whose compatibility yields the non-linear Einstein equation of interest.
This method applies whenever one is seeking a space-time with a sufficient number of commuting and hypersurface orthogonal Killing vectors. For $D=4$ one can use inverse scattering to construct stationary and axisymmetric solutions
(two Killing vectors), for $D=5$ one requires an additional space-like Killing vector to render the system integrable in the inverse scattering sense. The power of the inverse scattering method is that the construction is reduced to
a purely algebraic problem for the data entering the solitonic ansatz for a solution of the linear system~\cite{BZ1, BZ2}.

There are many other gravitational systems with matter to which one would like to apply the inverse scattering method. A number of examples can be constructed from string theory where one is led to supergravity theories and
the solutions sought include charged black holes. The class of models considered typically involves a finite-dimensional symmetry group $\mathrm{G}$ that acts as a solution generating group on the three-dimensional reduced system
(one Killing vector less than for the inverse scattering method). For pure $D=4$, this group is Ehlers's $\mathrm{SL}(2,\mathbb{R})$~\cite{Ehlers} while for maximal supergravity it is
$\mathrm{E}_{8(8)}$~\cite{Cremmer:1979up, Marcus:1983hb}. A list of all such three-dimensional gravity-matter models with symmetry $\mathrm{G}$ can be found in~\cite{Breitenlohner:1987dg}.
Unfortunately, the method of inverse scattering as developed in~\cite{BZ1, BZ2} is not directly applicable to all these cases since the soliton ansatz does not necessarily respect the structure of the group $\mathrm{G}$;
see for example the discussion in~\cite{FJRV} for the case $\mathrm{G}=\mathrm{G}_{2(2)}$ that arises for minimal $D=5$ supergravity.

Long ago, Breitenlohner and Maison (BM) have constructed a linear system that is different from that of Belinski and Zakharov (BZ) and that takes the structure of $\mathrm{G}$ into account~\cite{BM}.
The relation between the two linear systems was studied in~\cite{BM, FJRV, KKV}. The BM linear system has not been used extensively for solution generation although in~\cite{BMnotes} it was shown how to
implement a BZ like inverse scattering for $\mathrm{SL}(n,\reals)$. It is the purpose of the present article to describe how to use the BM linear system to generate solutions for more general groups $\mathrm{G}$.
We will focus mainly on the case $\mathrm{G}=\mathrm{SO}(4,4)$ for concreteness.  $\mathrm{G}=\mathrm{SO}(4,4)$ is the symmetry that is relevant for the STU model that has multiple constructions from string theory and
whose solutions have attracted a lot of attention over the years~\cite{STU1, STU2, STU3, STU4, Chow:2013tia}. Our methods do, however, apply more generally and we make some remarks in that direction at the end of the paper.

For the standard BZ inverse scattering method one constructs a generating function that has simple poles in the spectral parameters and the residues at these poles are of rank one. A major difference that arises for more
general groups is that the rank of the residue can be larger and therefore one needs to associate more data with any given pole. We will show this explicitly for $\mathrm{G}=\mathrm{SO}(4,4)$ where the rank is two and present
a general formalism in section~\ref{sec:rankR}. As a model example of our formalism we show how to recover the four-charge Cveti\v{c}-Youm solution~\cite{CY, STU3}.

The structure of this article is as follows. In section~\ref{sec:prelim} we establish our conventions and review the BM linear system. In section~\ref{sec:RHfac}, we demonstrate how to solve the linear system
for $\mathrm{G}=\mathrm{SO}(4,4)$  case with rank two residues in general and work out the Cveti\v{c}--Youm solution as a detailed example in section~\ref{sec:4charge}. Section~\ref{sec:rankR} contains the general formalism
for other groups and general ranks and concluding remarks can be found in section~\ref{sec:concl}. Appendix~\ref{app:conv} contains some more technical details on our choice of parametrization of $\mathrm{SO}(4,4)$ in terms
of the physical fields and appendix~\ref{app:Scalars} contains the explicit expression for the scalar fields for the four-charge black hole.

\section{Preliminaries: Lagrangian and linear system}
\label{sec:prelim}

\subsection{The three-dimensional system}
\label{sec:3dsystem}

We assume that there is a three-dimensional gravity-matter system that has a global symmetry group $\mathrm{G}$ and a local symmetry group $\mathrm{K}$ that is maximal subgroup of $\mathrm{G}$.
The elements $k\in \mathrm{K}$ satisfy $k^\# k =1$, where the `hash' denotes some generalized anti-involution. For $\mathrm{G}=\mathrm{SL}(n,\reals)$ and $\mathrm{K}=\mathrm{SO}(n)$ this operation is
just the usual transposition $k^\#=k^T$ but it can be different in general.

The three-dimensional system is given by\footnote{\label{normfn}We have changed the normalization of the scalar $\mathrm{G}/\mathrm{K}$ sector by a factor of $1/2$ compared to~\cite{KKV}.}
\begin{align}
\label{3dimsys}
\mathcal{L}_3 = \sqrt{g_3} \left( R_3 -\frac12 g^{\mu\nu}\mathrm{Tr}(P_\mu P_\nu) \right),
\end{align}
where $P_\mu$ is determined by $V\in \mathrm{G}/\mathrm{K}$ through
\begin{align}
P_\mu = \frac12\left(\partial_\mu V\cdot V^{-1} + ( \partial_\mu V\cdot V^{-1})^\# \right).
\end{align}

This system has the required symmetries that act on $V$ by
\begin{align}
\label{Gtrm}
V(x) \to k(x) V(x) g,
\end{align}
with a global $g\in \mathrm{G}$ and a local gauge transformation $k(x)\in \mathrm{K}$. A convenient object is the $x$-dependent
\begin{align}
M(x) = V^\#(x) V (x)\quad\quad
\textrm{with}\quad
M(x) \to g^\# M(x) g,
\end{align}
and that is thus independent of the choice of gauge.

\subsection{STU gravity}

The $D=4$ STU model fits into this picture when one considers stationary solutions. In this case $\mathrm{G}=\mathrm{SO}(4,4)$ and $\mathrm{K}=\mathrm{SO}(2,2)\times \mathrm{SO}(2,2)$~\cite{Breitenlohner:1987dg}.
The operation $\#$ can be given a more explicit expression if one chooses to represent the scalars $V\in \mathrm{G}/\mathrm{K}$ as $(8\times 8)$-matrices that leave invariant the metric
\begin{align}
\eta = \begin{pmatrix}
0_4 & \id_4\\
\id_4&0_4
\end{pmatrix},
\end{align}
that is written in block form with unit and zero matrices. Matrices $g$ satisfying $g^T \eta g = \eta$ belong to $\mathrm{SO}(4,4)$.
The subgroup $\mathrm{K}=\mathrm{SO}(2,2)\times \mathrm{SO}(2,2)$ then satisfies the further constraint that it leaves invariant~\cite{Bossard:2009we}
\begin{align}
\label{etaprimed}
\eta' = \mathrm{diag}(-1,-1,1,1,-1,-1,1,1),
\end{align}
and we have $V^\#= \eta' V^T \eta'$.

\subsection{Two-dimensional reduction and BM Linear system}

Following the discussion in \cite{Breitenlohner:1987dg, BM}, we consider further reduction of the system (\ref{3dimsys}) over the spacelike Killing vector $\partial_\varphi$, thereby obtaining an effectively two-dimensional system.
The three-dimensional metric can be written as
\begin{align}
\label{32met}
 ds_{3}^2=f^2 ds_2^{2}+\rho^2 d\varphi^2\,,
\end{align}
where the function $f$ multiplying the two-dimensional metric is called the conformal factor. Choosing Weyl coordinates $x^m=(\rho,z)$, the flat two-dimensional base metric is $ds^2_2=d\rho^2+dz^2$.
The equations of motion of the two-dimensional system read
\begin{subequations}
\label{eqns2d}
\begin{align}
 \pm if^{-1}\partial_{\pm}f=\frac{\rho}{4}\mathrm{Tr}\left(P_{\pm}P_{\pm}\right),\label{confeqn}\\
 D_m\left(\rho P^{m}\right)=0\label{sigmaeqn},
\end{align}
\end{subequations}
where we used the ``light-cone'' coordinates $ x^{\pm}=\frac{1}{2}(z\mp i \rho)$ to simplify the form of the equations.
Given a solution of (\ref{sigmaeqn}),
the function $f$ is obtained simply by integrating equation (\ref{confeqn}). Therefore, developing a strategy to obtain solutions is mostly concentrated on equation (\ref{sigmaeqn}).
In fact, this equation is shown to be integrable and
can be represented by a Lax pair or linear system. This means that there exists a system of linear equations whose compatibility condition is exactly the non-linear equation we wish to solve.
The functions we solve for
in the linear system depend on an additional parameter $t$, called the spectral parameter.

We define the generalized coset element $\cV(t,x)$, that has the form (suppressing the $x$-dependence)
\begin{align}
\label{cVexpansion}
 \cV(t)=V_0+tV_1+\frac{1}{2}t^2 V_2+...\,,
\end{align}
such that
\begin{align}
 \lim_{t \to 0}\cV(t)=V_0:=V,
\end{align}
and is a regular function in $t$ around $t=0$. The linear equations, referred to as the Breitenlohner-Maison (BM) linear system~\cite{BM,Nicolai:1991tt}
\begin{align}
\label{linsys}
 \partial_\pm \cV \cV^{-1}=\frac{1\mp it}{1\pm it}P_\pm +Q_\pm,
\end{align}
can be viewed as the generalisation of the relation $\partial_\pm V V^{-1}=P_\pm +Q_\pm$ for the Lie algebra-valued expression $\partial_\pm VV^{-1}$, in light of the Lie algebra decomposition
under the symmetric space automorphism. The integrability condition
\begin{align}
 \partial_+\left(\partial_{-}\cV\cV^{-1}\right)-\partial_{-}\left(\partial_{+}\cV\cV^{-1}\right)-\left[\partial_{+}\cV\cV^{-1},\partial_{-}\cV\cV^{-1}\right]=0,
\end{align}
yields the equation (\ref{sigmaeqn}) with the additional requirement that $t$ be a function which satisfies the differential equation
\begin{align}
 t^{-1}\partial_\pm t=\frac{1\mp it}{1\pm it}\rho^{-1}\partial_\pm \rho\,.
\end{align}
Integrating this equation, leads to a quadratic equation for $t$ with solutions
\begin{align}
\label{tbranches}
 t_\pm=\frac{1}{\rho}\left[(z-w)\pm\sqrt{(z-w)^2 +\rho^2}\right].
\end{align}
The integration constant $w$ can be regarded as an alternative, $x$-independent spectral parameter. Equation (\ref{tbranches}) defines a two-sheeted Riemann surface over the complex $w$-plane.
We choose the solution with the plus sign as the physical sheet and have $t$ to mean $t_+$ hereafter.

The existence of the linear system  (\ref{linsys}) that equivalently poses the problem at hand, exhibits not only that the two-dimensional
gravity system is integrable, but reveals
its symmetry properties as well. The generalized coset element $\cV(t,x)$, transforms under an enlarged symmetry group as
\begin{align}
\label{cVtransf}
 \cV(t) \rightarrow k(t)\cV(t)g(w)\,,
\end{align}
in a manner analogous to the gauge-preserving transfomations (\ref{Gtrm}) of $V\in \mathrm{G}/\mathrm{K}$. The general global transformation $g$ has now a dependence on the constant spectral parameter $w$ and $k(t)$
is the local compensating
transformation that brings $\cV$ back to the form (\ref{cVexpansion}). The subset of maps $g(w)$ from $S^1 \subset\mathbb{C}$ into $\mathrm{G}$ constitute the loop group $\hat{\mathrm{G}}$. This already shows that
the symmetry group
of the two-dimensional system includes the infinite-dimensional loop group associated to the finite group $\mathrm{G}$. In fact, the group of transformations involves the full affine extension of $\mathrm{G}$,
which comprises the central extension acting on the conformal factor $f$~\cite{BM}.

The symmetric space automorphism $\#$ admits a generalization for the enlarged symmetry group and its action on the functions $\cV(t)$ is given by

\begin{align}
 \left(\cV(t)\right)^{\#}=\cV^{\#}\left(-\frac{1}{t}\right).
\end{align}
With this definition, it can be shown that for any solution $\cV$ of (\ref{linsys}) the quantity $\partial_\pm \cV\cV^{-1}$ is anti-invariant under the $\#$ -involution induced on the associated Lie algebra.
This means that if $\cV(t)$ is a solution of (\ref{linsys}),
then the function $\left(\cV(t)\right)^{\#}$ is also a (generally distinct) solution.

In principle, given a seed solution $\cV(t)$ one could obtain new solutions $\cV^{g}(t)$ through the transformation \eqref{cVtransf}. However, in this approach one needs to determine $k(t)$, a task that is generally
quite hard. Alternatively, we can construct a function, analogous to $M=V^{\#}V$, called the monodromy matrix
\begin{align}
\label{RHp}
\cM(w)=\left(\cV(t)\right)^{\#}\cV(t)=\cV^{\#}\left(-\frac{1}{t}\right)\cV(t),
\end{align}
which transforms as
\begin{align}
\label{cMtrm}
 \cM(w)\rightarrow \cM^{g}(w):=g^{\#}(w)\cM(w)g(w)\,,
\end{align}
thus evading knowledge of the element $k(t)$. The $\#$-properties of (\ref{linsys}) imply that $\cM(w)$ is constant: $\partial_\pm \cM(w)=0$. Solutions can be now obtained from the factorization of $\cM^{g}(w)$ into
$\left(\cV^{g}(t)\right)^{\#}\cV^{g}(t)$. This is a Riemann--Hilbert problem, that is generally difficult to solve. However, in special circumstances, it becomes a purely algebraic procedure, as described in the following section.
Generally, the physical fields can be obtained from $\cV^{g}(t)$ by taking the limit $t\to 0$. On top of the solution of the Riemann--Hilbert problem (\ref{RHp}) we also need to determine the conformal factor $f$ by integrating
(\ref{confeqn}). In the algebraic case considered in the next section this is also easy to accomplish.

As in our previous work \cite{KKV}, in this article we always work with flat space
\be
\cV(t) = \id \qquad \mbox{and} \qquad f =1,
\ee as seed solution. Thus, from now on we simply drop the superscript $g$ from $\cM^{g}(w)$ and $\cV^{g}$ and think of being given a monodromy matrix $\cM(w)$ that needs to be factorized to find $\cV(t)$.

\section{Riemann--Hilbert factorization for $\mathrm{SO}(4,4)$}
\label{sec:RHfac}

We construct the monodromy matrix $\cM$ as
\begin{align}
 \cM=\cV^{\#}\left(-\frac{1}{t},x\right) \cV(t,x)= \h' \cV^{T}\left(-\frac{1}{t},x\right)\h'\cV(t,x),
\end{align}
where $\h'$ is the quadratic form of (\ref{etaprimed}) preserved by $\mathrm{SO}(2,2)\times \mathrm{SO}(2,2)$ and
\be
g^\#=\h'g^{T}\h'^{-1},\quad \forall \quad  g \in \mathrm{SO}(4,4).
\ee The matrix $\cM$ is by
construction an element in $\mathrm{SO}(4,4)$  (as $\cV \in \mathrm{SO}(4,4)$). As mentioned in the previous section, involution symmetry together with the Lax equations imply that
$\partial_\mu \cM = 0$, i.e., $\cM$ is independent of the spacetime coordinates $(\rho,z)$ and is a function of $w$ alone \cite{BM, Nicolai:1991tt}. Since $w$ is invariant under $t\rightarrow -1/t$,
it follows that $\cM$ is also invariant under simultaneous action of the generalized transposition $\#$ and the exchange $t \to -1/t$:
\begin{align}
 \cM^\#=\h'\cV^{T}(t,x)\h'\cV\left(-\frac{1}{t},x\right)= \cM.
\end{align}

In order to find $\cV(t)$ from $\cM$, we wish to factorize the matrix $\cM$ in the form
\be
\cM(w)=A_{-}^{\#}(t,x)M(x)A_{+}(t,x)
\ee
with $A_+(t)$ containing only positive powers of $t$~\cite{BM,BMnotes} and where the matrices $A_{\pm}$ satisfy the relation \cite{BMnotes, KKV}
\be
A_{-}(t,x)=A_{+}\left(-\frac{1}{t},x\right),
\ee
and $M^{\#}(x)=M(x)$.  We also require matrices $A_{\pm}(t,x)$ to be in $\mathrm{SO}(4,4)$. Furthermore we factorize $M(x) = V^{\#}(x) V(x)$ so that
\be
\cV(t,x) = V(x)A_{+}(t,x).
\ee

\subsection{Solution of the Riemann--Hilbert problem}

We restrict ourselves to the class of matrices $\cM(w)$ that have $N$ simple poles at locations $w=w_k$ that can be expressed in the form,
\begin{subequations}
\bea
\cM(w) &=& \id + \sum_{k=1}^N \frac{A_k}{w-w_k}, \\
\cM^{-1}(w) &=& \h \cM^{T}\h=\h\left(\id + \sum_{k=1}^N \frac{A^{T}_k}{w-w_k}\right)\h.
\eea
\end{subequations}
The matrix $\h$ is the quadratic form preserved by $\mathrm{SO}(4,4)$.

Unlike the case of  $\mathrm{SL}(n,\RR)$ considered in \cite{BMnotes, KKV} where the residue matrices $A_k$ are  taken to be of rank one, in the present analysis we take the residue matrices $A_k$ to be
of \emph{rank two}. In the following, in particular in the next section, it will become clear that the rank-two case corresponds to the simple solutions of physical interest. An intuitive way
to appreciate this is via the restriction of the general $\mathrm{SO}(4,4)$ matrix $M(x)$ to four-dimensional vacuum gravity. The structure of the restricted matrix is such that the Ehlers SL(2) representative
of four-dimensional vacuum gravity enters \emph{two times}, suggesting that the residue matrices in $\cM(w)$ should be taken to be of rank two in order to connect to solutions of vacuum gravity. A related observation
was also made in \cite{FJRV}, where in the context of the BZ method it was pointed out that for minimal supergravity, soliton transformations must be applied \emph{in pairs} in order to preserve the coset structure.

Using the expression
 \be
  \frac{1}{w-w_k} = \nu_k \left( \frac{t_k}{t-t_k}+ \frac{1}{1+t t_k}\right),
 \ee
where $t_k$ is the value of (\ref{tbranches}) at $w=w_k$, and
\be
\nu_k = -\frac{2}{\rho\left(t_k + \frac1{t_k}\right)},
\ee
we can write
\begin{align}
\label{texpansion}
 \cM(t,x)=\id+\sum_{k=1}^N \frac{\nu_k t_k A_k}{t-t_k}+\sum_{k=1}^N \frac{\nu_k A_k}{1+tt_k}\,.
\end{align}
The residue matrices $A_k$ can be factorized and parameterized as follows,
\begin{align}
A_k =  \alpha_k a_k a_k^T \h'-\b_k (\h b_k)(\h b_k)^{T}\h',
\label{Ak}
\end{align}
where $a_k$ and $b_k$ are 8-dimensional constant vectors. At first sight this choice may not look transparent or obvious, but its advantages will become clear very soon. Note that by construction, the matrices $A_k$ \eqref{Ak} satisfy
\be
A^{\#}_k=A_k,
\ee
as they should, since $\cM(w)$ satisfies this property. In order to deduce properties of the vectors $a_k$ and $b_k$,
we study the pole structure of the product $\cM(t,x)\cM^{-1}(t,x)$ or equivalently the pole structure of $\cM(t,x)\h\cM^{T}(t,x)$.
The absence of double poles in this product at $t=-1/t_k$ implies the conditions
 \begin{align}
 \label{dpcond}
  A_k\h A^{T}_k=0\quad \text{for all}\,\,k\,.
 \end{align}
These conditions are fulfilled when the vectors satisfy the following relations,
\begin{subequations}
\label{vecconds}
 \bea
\label{aconds}
 a^{T}_k\h a_k&=&0,\\
 \label{bconds}
 b^{T}_k\h b_k&=&0,\\
 \label{aborthog}
 a^{T}_k b_k&=&0,
\eea
\end{subequations}
for all $k$. The absence of single poles in the product $\cM(t,x)\h\cM^{T}(t,x)$ at $t=-1/t_k$
results in the conditions
\be
\mathcal{A}_k \h A^{T}_k=-A_k \h \mathcal{A}^{T}_k,
\label{singlepole}
\ee
where matrices $\mathcal{A}_k$ are defined as
\begin{align}
\mathcal{A}_k=\left.\left(\cM(t,x)-\frac{\nu_k A_k}{1+tt_k}\right)\right |_{t \rightarrow -\frac{1}{t_k}}\,.
\end{align}
The condition \eqref{singlepole} explicitly reads
\begin{align}
\label{gammacond}
 \mathcal{A}_k\h \h' \a_k a_k a^{T}_k-\mathcal{A}_k \h \h' \b_k (\h b_k)(\h b_k)^{T}=-\a_k a_k a^{T}_k \h' \h \mathcal{A}^{T}_k+\b_k (\h b_k)(\h b_k)^{T}\h' \h \mathcal{A}^{T}_k,
\end{align}
which is satisfied if there exist numbers $\gamma_k$ such that
\begin{subequations}
\bea
\mathcal{A}_k\h \h' a_k &=& \nu_k \b_k \gamma_k (\h b_k), \label{gammaone} \\
(\h b_k)^{T}\h' \h \mathcal{A}^{T}_k &=& \nu_k \a_k \gamma_k a^{T}_k.  \label{gammatwo}
\eea
\end{subequations}

Recall that, in order to solve the Riemann--Hilbert problem, we wish to factorize the matrix $\cM$ in the form
\be
\cM(w)=A_{-}^{\#}(t,x)M(x)A_{+}(t,x)
\ee
with matrices $A_{\pm}$ satisfying the relation
\be
A_{-}(t,x)=A_{+}\left(-\frac{1}{t},x\right),
\ee
and $M^{\#}(x)=M(x)$.  We also require matrices $A_{\pm}(t,x)$ to be matrices in $\mathrm{SO}(4,4)$. Furthermore we factorize $M(x) = V^{\#}(x) V(x)$ so that
\be
\cV(t,x) = V(x)A_{+}(t,x).
\ee
The analyticity properties (\ref{cVexpansion}) of the resulting $\cV(t,x)$ in the neighbourhood of $t=0$ require that
the poles at $t=-1/t_k$ come from the factor $A_{+}$~\cite{BM,BMnotes}. We therefore make the ans\"atze generalizing the ones used in \cite{BMnotes, KKV}
 \be
   A_{+}(t)=\id-\sum_{k=1}^N \frac{t C_k}{1+tt_k}\,,
\ee
with the parametrization of matrices $C_k$ as follows
\be
 C_k=c_k a^{T}_k \h'-(\h d_k) (\h b_k)^{T}\h'\,.
\ee
As in the  $\mathrm{SL}(n,\RR)$ case, the vectors $a_k, b_k, c_k$, and $d_k$ are not all independent and determining their relation amounts to solving the Riemann--Hilbert problem.

In order to determine the vectors $c_k$ and $d_k$ we study the poles in the product $A_{+}(t)\h\cM^{T}(t,x)$ at $t=-1/t_k$. The condition for no double poles is
\begin{align}
 C_k \h A^{T}_k=0\,,
\end{align}
which is fulfilled when the conditions \eqref{vecconds} hold. Furthermore, we need to ensure that the product  $A_{+}(t)\h\cM^{T}(t,x)$ has no single poles at $t=-1/t_k$. This requirement is equivalent to
\be
\label{condAplus}
 t^{-1}_k C_k \h \mathcal{A}^{T}_k+\left.\left(A_{+}+\frac{t C_k}{1+tt_k}\right)\right|_{t=-\frac{1}{t_k}} \h \nu_k A^{T}_k=0.
\ee
Writing equation (\ref{condAplus}) in terms of the vectors $a_k, b_k, c_k$, and $d_k$ and using relations \eqref{gammaone} and \eqref{gammatwo}, we arrive at
\begin{align}
 t^{-1}_k\left(c_k \nu_k \b_k \gamma_k (\h b_k)^{T}-(\h d_k) \nu_k\a_k\gamma_k a^{T}_k\right)+\nu_k \a_k\h \h' a_k a^{T}_k-\nu_k\b_k\h\h'(\h b_k)(\h b_k)^{T}\no\\
 +\sum_{l=1 \atop l\neq k}^N\frac{1}{t_k-t_l}\left(c_l a^{T}_l\h'-(\h d_l) (\h b_l)^{T}\h'\right)\h \nu_k\left(\h'\a_k a_k a^{T}_k-\h'\b_k(\h b_k)(\h b_k)^{T}\right)=0.
\end{align}
This condition is satisfied when the following two conditions are satisfied
\be
 -t^{-1}_k (\h d_k) \nu_k\a_k\gamma_k +\nu_k \a_k\h \h' a_k+\sum_{l=1 \atop l\neq k}^N\frac{\nu_k \a_k}{t_k-t_l}\left(c_l a^{T}_l\h a_k -(\h d_l) (\h b_l)^{T}\h a_k\right)=0,
 \label{messy1}
\ee
and
\be
 t^{-1}_k c_k \nu_k\b_k\gamma_k -\nu_k \b_k\h \h'(\h b_k)-\sum_{l=1 \atop l\neq k}^N\frac{\nu_k \b_k}{t_k-t_l}\left(c_l a^{T}_l\h (\h b_k) -(\h d_l) (\h b_l)^{T}\h (\h b_k)\right)=0.
 \label{messy2}
\ee
Assuming furthermore that the vectors $a_k$, $b_k$ satisfy
\begin{subequations}
\label{blockvectors}
\bea
a^{T}_l \h a_k &=&0, \\
b^{T}_l \h b_k &=&0,
\eea
\end{subequations}
for $l\neq k$, then the relations \eqref{messy1} and \eqref{messy2} simplify to
\bea
\h' a_k &=& \frac{\g_k}{t_k} d_k+\sum_{l\neq k}^N\frac{1}{t_k-t_l}d_l\left(a^{T}_k b_l\right),\\
\h' b_k &=& \frac{\g_k}{t_k} c_k +\sum_{l\neq k}^N \frac{1}{t_l-t_k} c_l \left(a^{T}_l b_k\right).
\eea
These relations can be written as matrix equations
\begin{subequations}
\begin{align}
\label{dmat}
 \h' a = d\, \G^{T},\\
 \label{cmat}
 \h' b= c\, \G\,,
\end{align}
\end{subequations}
where $a, b, c$, and $d$ are $8\times N$ matrices whose columns are the vectors $a_k, b_k, c_k$, $d_k$ respectively and $\G$ is a $N \times N$ matrix with elements
\begin{align}
 \label{Gdef}
\G_{kl} = \left\{ \begin{array}{ll}
 \frac{\gamma_k}{t_k} &\mbox{\qquad for \qquad $k=l$} \\
  \frac{a_k^T b_l}{t_k-t_l}  &\mbox{\qquad for \qquad $k \neq l$.}
       \end{array} \right.
\end{align}
Solving equations \eqref{dmat} and \eqref{cmat} for $c$ and $d$ we find the matrix $A_{+}(t,x)$ as
\begin{align}
\label{Aplusmat}
 A_{+}(t)=\id-\h'b\G^{-1}\frac{t}{\id+tT}a^{T}\h'+\h\h'a \left(\G^{T}\right)^{-1}\frac{t}{\id+tT}b^{T}\h\h',
\end{align}
where to avoid notational clutter we use $T$ to denote the $N\times N$ diagonal matrix with entries $t_k$. Taking the limit of the inverse of \eqref{Aplusmat} as $t\rightarrow\infty$ we get the matrix $M(x)$,
\be
M(x) = A^{-1}_{+}(\infty)=\h A^{T}_{+}(\infty)\h,
\label{matrixM}
\ee
with
\begin{align}
\label{AplusT}
 A^{T}_{+}(\infty)= \id-\h' a T^{-1} \left(\G^{-1}\right)^{T} b^{T}\h'+\h'\h b T^{-1}\G^{-1} a^{T}\h'\h.
\end{align}

If we furthermore assume that $a^{T}_l  b_k=-a^{T}_k  b_l$ for $l\neq k$, i.e., that the $\G$ matrix is symmetric, then  expression (\ref{AplusT}) becomes
\be
 A^{T}_{+}(\infty)= \id-\h' a T^{-1} \G^{-1} b^{T}\h'+\h'\h b T^{-1}\G^{-1} a^{T}\h'\h.
\label{main}
\ee

In the next section, we see that all assumptions made in the above analysis are satisfied for the four-charge black holes --- one of most studied set-up in four-dimensional STU supergravity.
We believe that various assumptions made above are also satisfied in more general settings of physical interest.

\subsection{Computation of the conformal factor}
\label{app:CF}

The conformal factor is determined by integration of equation (\ref{confeqn}). This proceeds exactly along the same lines as in appendix~A of~\cite{KKV}, keeping in mind the change of normalization
of the scalars, cf.~footnote~\ref{normfn}. We do not repeat all the steps here but only indicate a few intermediate results where the rank-two property of the residues enters.

For evaluating (\ref{confeqn}) we need to detemine $\mathrm{Tr}(P_\pm P_\pm)$. This is most conveniently done in terms of evaluating first $A_+^{-1}(t) \frac{\partial}{\partial t} A_+(t)$~\cite{BMnotes,KKV}. For the value of $A_+(t)$ determined in (\ref{Aplusmat}) one finds
\be
A_{+}^{-1}(t) \frac{\partial}{\partial t}  A_{+}(t) =
- \eta' b\frac{\id}{\id + t T} \Gamma^{-1} \frac{\id}{\id + t T} a^T \eta'
+ \eta \eta' a \frac{\id}{\id + t T} \Gamma^{-1} \frac{\id}{\id + t T} b^T \eta \eta',
\ee
which is now composed of two terms reflecting the rank-two nature of the residues. The next important intermediate quantity is
\be
\mathrm{Tr} (A_{+}^{-1}(\pm i) \dot A_{+}(\pm i))^2 = 2 \sum_{k,l,m,n}\frac{\Gamma^{-1}_{kl}\Gamma^{-1}_{mn}}{(1 \pm i t_k)(1 \pm i t_l)(1 \pm i t_m)(1 \pm i t_n)}\mathrm{Tr} (b_k a_l^T b_m a_n^T),
\ee
where the factor of $2$ is due to the increased rank. Otherwise the result is exactly equal to the one in~\cite{KKV}. The changed normalization of the scalars cancels this factor of $2$ so that we obtain the conformal factor as
\be
\label{CF44}
f^2 = k_{\rom{BM}} \cdot \prod_{k=1}^{N} (t_k \nu_k) \cdot \det \Gamma,
\ee
where $k_{\rom{BM}}$ is an integration constant.

\section{Construction of the four-charge black hole}
\label{sec:4charge}
In this section we present a fairly non-trivial implementation of the inverse scattering method of the previous section. We construct the four-charge black hole of  STU supergravity from flat space.
This construction illustrates all the steps of the algorithm presented earlier.

As in the $\mathrm{SL}(n,\RR)$ case studied in \cite{BMnotes, KKV} the main difficulty in constructing the general multisoliton solutions using the BM method lies in finding the appropriate meromorphic matrices $\cM(w)$
that satisfy the various requirements of the previous section and satisfy the coset constraints.  It turns out that in the two-soliton case, as in the $\mathrm{SL}(n,\RR)$ models, finding appropriate solitonic matrices is not
difficult. We start with monodromy matrices of the form
\be
\cM(w) = \id + \frac{A_1}{w-c} + \frac{A_2}{w+c},
\label{cMw2}
\ee
where
\begin{subequations}
\bea
A_1 &=& \alpha_1 a_1 a_1^T \eta' - \beta_1 (\eta b_1)(\eta b_1)^T \eta', \\
A_2 &=& \alpha_2 a_2 a_2^T \eta' - \beta_2 (\eta b_2)(\eta b_2)^T \eta',
\eea
\end{subequations}
and where $a_1, a_2$ and $b_1, b_2$ are 8-dimensional vectors. In writing \eqref{cMw2} the location of the poles is chosen to be at $w_1=+c$ and $w_2=-c$. This choice can always be made by `shifting' the axis (see \cite{KKV} for a more detailed discussion on this). For finding the vectors $a_1, a_2$ and $b_1, b_2$ corresponding to  the four-charge black hole, let us start by looking at corresponding vectors for the Kerr-black hole in the $\mathrm{SO}(4,4)$ context. Analyzing the structure of the $\mathrm{SO}(4,4)$ matrix  $M(x)$ and embedding of the Ehlers's $\mathrm{SL}(2,\RR)$ in it, we make the inspired ansatz for the $a$-vectors
\begin{subequations}
\bea
a_1 &=& (0,0,-\zeta ,0,0,0,0,1)^T, \\
a_2 &=& (0,0,1,0,0,0,0,-\zeta)^T.
\eea
\end{subequations}
Next we follow an algorithm similar to the one used in \cite{BMnotes, KKV} to construct the $b$-vectors. We first construct the matrix
$
a = (a_1, a_2),
$
next we find the $2\times2$ matrix $\xi = a^T \eta' a$ and choose
\be
b = (\sqrt{\det \xi}) \eta' a \xi^{-1} \epsilon \qquad \qquad \mbox{with} \qquad \qquad \epsilon  =
\left(
\begin{array}{cc}
0 & -1 \\
1 & 0
\end{array}
\right).
\ee
This results in $b$-vectors
\begin{subequations}
\bea
b_1 &=& (0,0,1,0,0,0,0,\zeta)^T \\
b_2 &=& (0,0,-\zeta,0,0,0,0,-1)^T.
\eea
\end{subequations}
Finally we must choose
\begin{align}
\alpha_1 &= +2 c \frac{1+\zeta ^2}{(1-\zeta ^2)^2},&
\alpha_2 &= -2 c \frac{1+\zeta ^2}{(1-\zeta ^2)^2},& \\
\beta_1  &= -2 c \frac{1+\zeta ^2}{(1-\zeta ^2)^2},&
\beta_2  &= +2 c \frac{1+\zeta ^2}{(1-\zeta ^2)^2},&
\end{align}
in order to satisfy the coset constraints. It can be readily verified that all the conditions required on the vectors from the previous section are satisfied in this construction. In particular we note that
\begin{subequations}
\begin{align}
&a_1^T \eta a_1 = 0,& & a_2^T \eta a_2 = 0,& & a_1^T \eta a_2 = 0, \label{rel1} \\
&b_1^T \eta b_1 = 0,& & b_2^T \eta b_2 = 0,& & b_1^T \eta b_2 = 0, \label{rel2} \\
&a_1^T b_1      = 0,& & a_2^T b_2      =0, & & a_1^T b_2 = -a_2^T b_1 = -1 +\zeta^2.  \label{rel3}
\end{align}
\end{subequations}
The above data results in the following matrix,
\be
\cM(w) =
\small \left(
\begin{array}{cccccccc}
 1 & 0 & 0 & 0 & 0 & 0 & 0 & 0 \\
 0 & 1 & 0 & 0 & 0 & 0 & 0 & 0 \\
 0 & 0 & 1 + \frac{2 m (m-w)}{w^2-c^2} & 0 & 0 & 0 & 0 & \frac{2 a m}{w^2-c^2} \\
 0 & 0 & 0 & 1+\frac{2 m (m-w)}{w^2-c^2} & 0 & 0 & -\frac{2 a m}{w^2-c^2} & 0 \\
 0 & 0 & 0 & 0 & 1 & 0 & 0 & 0 \\
 0 & 0 & 0 & 0 & 0 & 1 & 0 & 0 \\
 0 & 0 & 0 & -\frac{2 a m}{w^2-c^2} & 0 & 0 & 1+\frac{2 m (m+w)}{w^2-c^2} & 0 \\
 0 & 0 & \frac{2 a m}{w^2-c^2} & 0 & 0 & 0 & 0 & 1+\frac{2 m (m+w)}{w^2-c^2}
\end{array}
\right),
\normalsize
\ee
where (at some places) we have replaced $\zeta$ and $c$ in favor of $m$ and $a$. The relations between these  parameters are
\be
\zeta = \frac{c-m}{a}, \qquad \qquad c = \sqrt{m^2 -a^2}.
\ee
This matrix is precisely the $\mathrm{SO}(4,4)$ monodromy matrix for the Kerr metric -- factorization of it gives the Kerr-field.

Having obtained the monodromy matrix for the Kerr metric, generalization to the four-charge black hole is now straightforward. We simply conjugate the Kerr matrix with the appropriate group element,
\be
\cM_\rom{4-charge}(w) = g^\# \cM(w) g.
\ee
Since in our duality frame, the four-charge black hole corresponds to three-magnetic charges and one-electric charge, we act on  $\cM(w)$
with the following group element
\be
g= \exp[-\delta_0 (E_{q_0} + F_{q_0})]\cdot \exp[\delta_1 (E_{p^1} + F_{p^1})] \cdot \exp[\delta_2 (E_{p^2} + F_{p^2})] \cdot \exp[\delta_3 (E_{p^3} + F_{p^3})].
\ee
The transformed vectors are
\begin{subequations}
\label{trmvecs}
\bea
a_1 &=& (-c_0 s_1, - \zeta c_3 s_2, - \zeta c_2 c_3, - s_0 s_1, -c_1 s_0, - \zeta c_2 s_3, \zeta s_2 s_3, c_0 c_1)^T,\\
\label{vector1}
a_2 &=& (\zeta c_0 s_1, c_3 s_2, c_2 c_3, \zeta s_0 s_1, \zeta c_1 s_0, c_2 s_3, - s_2 s_3, - \zeta c_0 c_1)^T,\\
\label{vector2}
b_1 &=& (\zeta c_0 s_1,  -c_3 s_2, c_2 c_3, - \zeta s_0 s_1, \zeta c_1 s_0, - c_2 s_3, - s_2 s_3, \zeta c_0 c_1)^T,\\
\label{vector3}
b_2 &=& ( -c_0 s_1, \zeta c_3 s_2,  -\zeta c_2 c_3, s_0 s_1, - c_1 s_0, \zeta c_2 s_3, \zeta s_2 s_3, - c_0 c_1)^T,
\label{vector4}
\eea
\end{subequations}
where to avoid notational clutter we have introduced $c_i = \cosh \delta_i$ and $s_i = \sinh \delta_i$. Using these vectors we construct the monodromy matrix of the four-charge black hole. By group property it follows that relations \eqref{rel1}--\eqref{rel3} hold as it is. With these choices we find
\begin{subequations}
\bea
\gamma_1&=&\frac{2 \zeta  (1-\zeta^2) t_2 (1+t_1^2) }{(1+\zeta ^2)(t_1 - t_2)(1 + t_1 t_2)}, \label{gamma1} \\
\gamma_2&=&\frac{2 \zeta (1-\zeta^2)t_1 (1+t_2^2)  }{(1+\zeta ^2)(t_1 - t_2)(1 + t_1 t_2)}. \label{gamma2}
\eea
\end{subequations}
{}From these expressions we readily construct the $\Gamma$ matrix and using relations \eqref{dmat} and \eqref{cmat} we find the $c$ and $d$ vectors, and hence solve the factorization problem.
From expressions \eqref{matrixM} and \eqref{main} we find the final matrix $M(x)$ for the four-charge black hole.

The conformal factor, which is given by \eqref{CF44}, takes the form
\be
f^2= - 4 k_\rom{BM} t_1^2 t_2^2 (1-\zeta^2)^2 \frac{(1 + t_1 t_2)^2(1-\zeta^2)^2- 4 (t_1 - t_2)^2 \zeta^2}{(1 + t_1^2)(1+t_2^2)(t_1-t_2)^2(1+ t_1 t_2)^2(1+\zeta^2)^2 \rho^2}.
\label{conf4charge}
\ee
Using the conformal factor we construct the three-dimensional base metric. Using the base metric and the matrix $M(x)$, we can read off all physical fields.
Expressions for these fields are presented in appendix \ref{app:Scalars} along with some further details. In this way we recover the full set of fields for the four-charge black hole.

\section{Generalization of BM method: residues of rank~$r$}
\label{sec:rankR}

We now consider the general monodromy matrix
\begin{align}
 \cM(w)=\cV^{\#}\left(-\frac{1}{t},x\right)\cV(t,x)\,,
\end{align}
with $\cV(t,x)$ the generalization of $V(x)\in \mathrm{G}/\mathrm{K}$ that also depends on the spectral parameter $t$. The map $\#:\mathrm{G}\rightarrow \mathrm{G}$ is the anti-involution
already introduced in section~\ref{sec:3dsystem}.

For the $N$-soliton solution, one takes $\cM(w)$ to be a meromorphic function with $N$ simple poles at $w=w_k$ in the form:
\begin{align}
\label{cMgen}
 \cM(w) = \id + \sum_{k=1}^N \frac{A_k}{w-w_k},
\end{align}
and
\begin{align}
 \cM^{-1}(w) = \id - \sum_{k=1}^N \frac{B_k}{w-w_k}\,,
\end{align}
with $A_k, B_k$ the $x$-independent residue matrices.
The $t$-dependent expansions of $\cM$ read
\begin{align}
\label{texp}
 \cM(t,x)=\id+\sum_{k=1}^N \frac{\nu_k t_k A_k}{t-t_k}+\sum_{k=1}^N \frac{\nu_k A_k}{1+tt_k},
\end{align}
and
\begin{align}
 \cM^{-1}(t,x)=\id-\sum_{k=1}^N \frac{\nu_k t_k B_k}{t-t_k}-\sum_{k=1}^N \frac{\nu_k B_k}{1+tt_k}\,.
\end{align}
Let $A_k$, $B_k$ be diagonalizable matrices of size $n$ and rank $r$, $(r\leq n)$, which moreover satisfy $A_k=A_k^{\#} $ and $B_k=B_k^{\#}$. There exists a matrix $U_k$ satisfying $U_{k}^{-1}=U_{k}^{\#}$ and
a diagonal matrix $\Lambda_k$ such that
\begin{align}
 A_k=U_k \Lambda_k U_{k}^{\#}\,.
 \end{align}
Thus we can write the matrix $A_k$ (same treatment applies to $B_k$) in the form of a sum of rank one matrices as follows:
\begin{align}
\label{Afact}
 A_k=\sum_{\a=1}^{r} \lambda_{k}^{\a}u_k^{\a}v_k^{\a \,T}\,,
\end{align}
where $\lambda_{k}^{\a} $ are the non-zero entries of the diagonal matrix $\Lambda_k$. The vectors $u_k^{\a} $ and $v_{k}^{\a \,T}$ are the corresponding ($n$-dimensional) column vectors of matrix $U_k$ and
corresponding row vectors
of matrix $U_{k}^{\#}$ respectively.

One can write the previous rank one decomposition in a manifestly ``$\#$-invariant'' form when the action of the map $\#$ on $g\in \mathrm{G}$
is explicitly known (in the matrix representation of the group).
As an example consider the coset space $\mathrm{G}/\mathrm{K}=\mathrm{SO}(4,4)/\mathrm{SO}(2,2)\times \mathrm{SO}(2,2)$ with $\tau$ the involutive automorphism that fixes the subgroup
$\mathrm{SO}(2,2)\times \mathrm{SO}(2,2)$. The action of $\#$ on $g\in \mathrm{G}$ is given by
$g^{\#}=\eta'g^T \eta'$, with $\eta'$ the quadratic form preserved by $\mathrm{SO}(2,2)\times \mathrm{SO}(2,2)$. The residue matrices $A_k$ (similarly for $B_k$) can be expressed in the form
\begin{align}
 A_k=U_k \Lambda_k U_k^{\#}=U_k\eta'\Lambda_k \eta' \eta'U^{T}\eta'=U_k \Lambda '_{k}U^{T}\eta' 
 =\sum_{\a=1}^{r}\lambda_{k}^{'\a} u_{k}^{\a} u_{k}^{\a\, \#}\,,
\end{align}
 where we use the ``$\#$-invariance'' of the diagonal matrix $\Lambda_k$ and $\Lambda_{k}'=\eta'\Lambda_k$. Moreover, the $\#$ operation on column vectors is defined as $u_k^{\#}=u_k^{T}\eta' $
 and on row vectors as ${u_k^{T}}^{\#}=\eta' u_k $.
(Indeed, using this definition, we have that for any vector $v$ and a matrix $ S=vv^{\#}\in \mathrm{G}$, $ S^{\#}=S $).
Assuming we can adopt this notation in the general case and using the freedom to redefine the vectors and tune $\lambda_{k}^{\a}$ accordingly, one can write\footnote{The notation we have used earlier for the case
of $\mathrm{G}/\mathrm{K}=\mathrm{SO}(4,4)/\mathrm{SO}(2,2)\times \mathrm{SO}(2,2)$ is somewhat different. However,
the previous notation can be readily translated in the general notation used in this section by identifying
$
p_k^{1}=a_k,
p_k^{2}=-\eta b_k,
q_k^{1}=\eta' b_k,
q_k^{2}=\eta \eta' a_k,
\alpha_k^{1}=-\beta_k^{2}=\alpha_k,
\alpha_k^{2}=-\beta_k^{1}=-\beta_k,
r_k^{1}=c_k,
r_k^{2}=\eta d_k,
s_k^{1}=\eta'd_k,
s_k^{2}=-\eta\eta'c_k
$
(with $\alpha_k, \beta_k$ the constants in section~\ref{sec:RHfac}) and using the $\#$ operation on vectors as defined above. }
\begin{align}
 A_k=\a_k\sum_{\a=1}^{r}p^{\a}_k p^{\a\,\#}_{k},\qquad B_k=\b_k\sum_{\a=1}^{r}q^{\a}_k q^{\a\,\#}_{k},
\end{align}
with $p_k^{\a},q_k^{\a}$ the redefined $n$-dimensional vectors and $\a_k,\b_k$ are constant parameters, not to be confused with the greek upper indices. The latter enumerate the vectors
with respect to the rank of the residue matrix, while the lower indices denoted by $k,l,...$ are the soliton indices and take values in $\left\lbrace 1,2,...,N\right\rbrace $.

Studying the pole structure of the product $\cM(t,x) \cM^{-1}(t,x)$ at $t=-\frac{1}{t_k}$, one can infer the required conditions on the vectors  $p_k^{\a},q_k^{\a}$.
The condition for no double poles in the product
$\cM(t,x) \cM^{-1}(t,x)$ at $t=-\frac{1}{t_k}$ is fulfilled when
\begin{align}
\label{pqcond}
 p^{\a\,\#}_{k}q^{\b}_{k}=0,\quad\text{for all}\,\, k\,\,\,\text{and}\,\,\a=1,2,...,r\,,\quad \b=1,2,...,r\,.
\end{align}
Furthermore, the absence of single poles in $\cM(t,x) \cM^{-1}(t,x)$ at $t=-\frac{1}{t_k}$ requires the condition
\begin{align}
 \mathcal{A}_k B_k=A_k \mathcal{A}^{k},
\end{align}
to be satisfied, with

\be
\mathcal{A}_{k}=\left.\left(\cM(t,x)-\frac{\nu_k A_k}{1+tt_k}\right)\right |_{t \rightarrow -\frac{1}{t_k}}, \
\qquad \qquad
\mathcal{A}^{k}=\left.\left(\cM^{-1}(t,x)+\frac{\nu_k B_k}{1+tt_k}\right)\right |_{t \rightarrow -\frac{1}{t_k}}.
\ee

 The demand is met if there exist
$\g^{\a}_{k}$ numbers
such that
\begin{align}
 \mathcal{A}_k q^{\a}_{k}=\nu_k\a_k \g^{\a}_{k} p^{\a}_{k}\,,\qquad\qquad p^{\a\,\#}_{k}\mathcal{A}^{k}=\nu_k \b_k\g^{\a}_{k} q^{\a\,\#}_{k}\,,
\end{align}
for all $k=1,2,...,N$ and $\a=1,2,...,r$.

The solution of the Riemann--Hilbert problem amounts to the factorization of $\cM$, with the expansion (\ref{texp}), in the form
\begin{align}
 \cM(w)=A_{-}^{\#}(t,x)M(x)A_{+}(t,x),
\end{align}
with $A_{-}(t,x)=A_{+}(-\frac{1}{t},x)$ and $M^{\#}(x)=M(x)$. The poles at $t=-\frac{1}{t_k}$ come from the factor $A_{+}$ and so we assume this matrix to be of the form
\begin{align}
 A_{+}=\id-\sum_{k=1}^N \frac{t C_k}{1+tt_k},
\end{align}
and
\begin{align}
\label{invAplus}
 A_{+}^{-1}=\id+\sum_{k=1}^N \frac{t D_k}{1+tt_k},
\end{align}
with $C_k= \sum\limits_{\a=1}^{r}r^{\a}_{k} p^{\a\,\#}_{k}$ and $D_k= \sum\limits_{\a=1}^{r}q^{\a}_{k} s^{\a\,\#}_{k}$.
In order to determine the vectors $r_k^{\a}$, we study the pole structure of the product $A_{+}(t)\cM^{-1}(t,x)$ at $t=-\frac{1}{t_k}$.
The absence of double poles yields the condition
\begin{align}
 C_k B_k=0,
\end{align}
and is fulfilled when (\ref{pqcond}) holds. The condition for no single poles is
\begin{align}
 t^{-2}_k C_k \mathcal{A}^{k}=\left.\left(A_{+} +\frac{t C_k}{1+tt_k}\right)\right |_{t \rightarrow -\frac{1}{t_k}} B_k \nu_k t^{-1}_k,
\end{align}
and is satisfied when
\begin{align}
 q^{\a}_k=t^{-1}_k r^{\a}_k \g^{\a}_k+\sum_{l\neq k}^N \sum_{\b=1}^{r} \frac{1}{t_l-t_k}r^{\b}_{l}p^{\b\,\#}_{l} q^{\a }_{k},
 \end{align}
that is, when these $rN$ vector equations hold. We can express them in a more compact way, in the form\footnote{These vector equations can be represented by the matrix equation $q=r\,\G$,
where $q$ is the $n\times rN$ matrix whose columns are the vectors  $q^{1}_1,q^{1}_2,...,q_N^{1},q^{2}_1\,,q^{2}_2,...,q_N^{2},...,q_1^{r},q_2^{r},...,q_N^{r}$ and the matrix $r$
 is defined similarly (with columns the $r_k^{\a}$ vectors).}
\begin{align}
\label{qeqn}
 q_B=\sum_{A=1}^{rN}r_A \G_{AB}\,,
\end{align}
where the capital indices $A,B$ take values in $\left\lbrace 1,2,...,rN\right\rbrace$ and each value uniquely determines a pair of indices $(k,\a)$. This can be done for example through the relations
\begin{align}
 k= \left\{ \begin{array}{ll}
  A\,\text{mod}\,N&\mbox{\qquad if \qquad $A\,\, \text{mod}\,\, N>0$} \\
  N &\mbox{\qquad if \qquad $A\,\, \text{mod}\,\, N=0$},
       \end{array} \right.\qquad\qquad\qquad\qquad \a=1+\left[\frac{A-1}{N}\right]\,,
\end{align}
where $\left[\cdot\right]$ denotes the integer part (floor function).
The matrix $\G$ is defined as the $rN\times rN$ block matrix with entries
\begin{align}
 \G^{\a\b}_{kl} = \left\{ \begin{array}{ll}
 \frac{\gamma^{\a}_k}{t_k}\delta_{\a\b}  &\mbox{\qquad for \qquad $k=l$} \\
  \frac{p^{\a\,\#}_k q^{\b}_l}{t_k-t_l}  &\mbox{\qquad for \qquad $k \neq l$,}
       \end{array} \right.
\end{align}
where the upper indices denote the block entry and the lower indices the entries of each block. It is a symmetric matrix under the condition $p^{\a\,\#}_{k} q^{\b}_{l}=-p^{\b\,\#}_{l} q^{\a}_{k}$
for $ k\neq l$ and all $\a,\b$ in $\left\lbrace 1,2,...,r\right\rbrace$. Moreover, when the condition $p^{\a\,\#}_{k} q^{\b}_{l}=0$ for $k \neq l$ and $\a \neq \b$ holds,
the off-diagonal blocks of $\G$ vanish (this is the case in all examples we have worked with so far). Solving (\ref{qeqn}) for the vectors $r_B$ we find
\begin{align}
 r_B=\sum_{A=1}^{rN} q_A\left(\G^{-1}\right)_{AB}\,.
\end{align}
There is one more set of vectors that we need to determine and these are the $s_k^{\a}$ in (\ref{invAplus}). The requirement that $\left(\cM(t,x) A_{+}^{-1}\right)^{\#}$ have no poles at $t=-\frac{1}{t_k}$
is fulfilled when
\begin{align}
 p^{\a}_k=t^{-1}_k s^{\a}_k \g^{\a}_k+\sum_{l\neq k}^N \sum_{\b=1}^{r} \frac{1}{t_k-t_l}s^{\b}_{l}p^{\a\,\#}_{k} q^{\b}_{l}
\quad\Longleftrightarrow \quad p_A =\sum_{B=1}^{rN} \G_{AB} s_B
\end{align}
and the equation for the vectors $s_A$ is\footnote{The matrix equation is now $p=s\,\G^{T}$, where $p$, $s$ are $n\times rN$ matrices whose columns are the vectors $p_k^{\a}$,$s_k^{\a}$ respectively
and are defined similarly to matrices $q$ and $r$.}
\begin{align}
 s_A=\sum_{B=1}^{rN}=\left(\G^{-1}\right)_{AB}p_B\,.
\end{align}
Finally, the matrix $M(x)$ is obtained by
\begin{align}
 M=A^{-1}_{+} (\infty)=\id+\sum_{A,B=1}^{rN}q_A t_{A}^{-1}\left(\G^{-1}\right)_{AB}p_B^{\#}\,,
\end{align}
where $t_A=t_k^\a=t_k$ for all values of $\a$.

\subsection*{Conformal factor}
The formula for the conformal factor in the multisoliton case with residues of rank $r$ is given by
\begin{align}
\label{CFr}
 f^4=k_\rom{BM}\cdot \det\G\cdot \prod_{A=1}^{rN}\left(t_A \nu_A\right)\no\\
 =k_\rom{BM}\cdot \det\G\cdot \prod_{k=1}^{N}\left(t_k \nu_k\right)^{r}\,.
\end{align}
This follows by a straightforward application of the computation of appendix A of~\cite{KKV} since the expression for $M$ is formally the same except for the enlarged range for the indices of $\Gamma_{AB}$.
The power on $f$ on the left-hand side of (\ref{CFr}) is due the changed normalization mentioned in footnote~\ref{normfn}.

We note that (\ref{CFr}) is consistent with (\ref{CF44}) since in the discussion of section~\ref{sec:RHfac} the vectors were assumed to satisfy (\ref{blockvectors}). In that case the matrix $\Gamma_{AB}$ becomes
block diagonal with $r$ repeated blocks of the matrix $\Gamma_{kl}$. Then $\det(\Gamma_{AB}) = \left(\det(\Gamma_{kl})\right)^r$ and this leads to the agreement between (\ref{CF44}) and (\ref{CFr}) when one takes into
account the different powers on $f$.

\section{Discussion}
\label{sec:concl}

In this paper we studied the integrability of STU supergravity and
proposed an inverse scattering technique for
this theory. Our main interest in performing this analysis is to make
available solution generating techniques based on integrability for set-ups where the standard BZ construction is not applicable.
Our approach makes use of the Geroch group (affine symmetry) of the
dimensionally reduced STU theory. We concentrated on Geroch group
matrices with simple poles only -- the so-called soliton sector. The
main difference compared to the SL$(n,\RR)$ analysis presented in
\cite{BMnotes, KKV} is that in the present $\mathrm{SO}(4,4)$ case the
rank of the residue matrices is two -- as opposed to one -- for simple
solutions of physical interest. In view of further generalization (and
future applications) of this technique we also presented a
generalization to arbitrary group $\mathrm{G}$ incorporating residue
matrices of arbitrary rank $r$.

Comparing our solution generating technique to that based on the finite-dimensional $\mathrm{G}$-symmetry used by many authors, we find that it is nicely consistent.
A (charging) transformation by a global element $k\in \mathrm{K}\subset \mathrm{G}$ rotates the matrix $\cM(w)$ according to (\ref{cMtrm}). Since $k$ is $w$-independent
it does not affect the location of the poles $w_k$ but rotates the residue matrices $A_k$ in (\ref{cMgen}) also according to (\ref{cMtrm}).
This induces a rotation of the vectors arising in the factorization (\ref{Afact}) but only in such a way that the matrix $\Gamma_{AB}$ does not change and consequently the conformal factor (\ref{CFr}) is unchanged.
The action of the symmetry is then the same that one would have in the three-dimensional system (\ref{3dimsys}).

There are many ways in which our study can be extended. The next
natural step would be to understand five-dimensional asymptotically flat
boundary conditions from the Geroch group point of view. This requires changing the asymptotic behavior of $\mathcal{M}(w)$ for $w\to\infty$. Together
with the results of the present paper, this will allow us to construct the
5d charged rotating Cveti\v{c}-Youm \cite{CY5d} metric which
in turn will lead to an inverse scattering construction of the JMaRT
fuzzball \cite{JMaRT}. Such a construction is highly desirable as it
will naturally lead to ways to generalize the JMaRT fuzzball. Various
problems in relation to five-dimensional black rings will also become
accessible once we incorporate five-dimensional asymptotically flat
boundary conditions in our formalism. We hope to report on these
issues in the near future.

On the technical side there is another difficulty that needs to be
overcome before our construction can be applied in its full potential.
Recall that, in order to apply our formalism for the construction of
the four-charge black hole we used the group property to find the vectors
\eqref{trmvecs} starting from that of the Kerr black
hole. For this computation, group rotation is sufficient, but we
expect that in more complicated situations, in particular for
configurations involving three or more poles, one needs to develop
some other algorithmic techniques to find appropriate vectors. In this
regard, ideas from the interval structure \cite{Harmark:2004rm,
Hollands:2007aj, Chen:2010zu} of gravitational solutions can be useful, but at the
moment this remains an open challenging problem.

More generally, since the five-dimensional version of the STU theory
has Chern-Simons terms in its Lagrangian, we expect a very large
family of non-trivial bubbling -- fuzzball-like -- solutions~\cite{Mathur:2005zp} to be
within reach of our proposed formalism; see \cite{BW} for a recent
discussion on this point. Although we have taken a significant step
forward in attacking this problem in this paper, some further
technical developments are necessary before such sought after
geometries can be explicitly constructed.

\subsection*{Acknowledgements}
We would like to thank Geoffrey Comp\`ere for correspondence. AV would also like to thank Albert Einstein Institute, Golm and Institute of Mathematical Sciences, Chennai for their warm hospitality where part of this work was done.

\appendix

\section{Conventions}
\label{app:conv}

In this appendix we detail the conventions that we are using for the STU model.

\subsection{The $\mathrm{SO}(4,4)$ group and its subgroups}

We adopt the conventions of~\cite{Bossard:2009we,Virmani:2012kw}. Thus we have the set of $\mathrm{SO}(4,4)$ generators labelled by
\begin{align}
H_\Lambda,\quad
E_\Lambda,\quad
F_\Lambda, \quad
E_{q_\Lambda},\quad
F_{q_\Lambda},\quad
E_{p^\Lambda},\quad
F_{p^\Lambda}
\end{align}
for $\Lambda=0,1,2,3$.
The subgroup relevant to time-like reductions is $\mathrm{SO}(2,2)\times\mathrm{SO}(2,2)\approx \mathrm{SL}(2)^4$; it is generated by
\begin{align}
K_\Lambda = E_\Lambda-F_\Lambda,\quad K_{q_\Lambda} = E_{q_\Lambda}+F_{q_\Lambda},\quad K_{p^\Lambda} = E_{p^\Lambda}+F_{p^\Lambda}.
\end{align}
The four commuting sets of  $\mathrm{SL}(2)$ generators in standard basis are for example given by
\allowdisplaybreaks{
\begin{subequations}
\label{sl24basis}
\begin{align}
h_0 &= \frac12 \left(-K_{q_0} + K_{p^1} + K_{p^2} + K_{p^3}\right),\\
h_1 &= \frac12 \left(+K_{q_0} - K_{p^1} + K_{p^2} + K_{p^3}\right),\\
h_2 &= \frac12 \left(+K_{q_0} + K_{p^1} - K_{p^2} + K_{p^3}\right),\\
h_3 &= \frac12 \left(+K_{q_0} + K_{p^1} + K_{p^2} - K_{p^3}\right),\\
e_0 &= \frac14 \left(-K_0 + K_1+K_2 + K_3 + K_{q_1} + K_{q_2} + K_{q_3} + K_{p^0}\right),\\
f_0 &= \frac14 \left(+K_0 - K_1-K_2 - K_3 + K_{q_1} + K_{q_2} + K_{q_3} + K_{p^0}\right),\\
e_1 &= \frac14 \left(+K_0 - K_1+K_2 + K_3 + K_{q_1} - K_{q_2} - K_{q_3} + K_{p^0}\right),\\
f_1 &= \frac14 \left(-K_0 - K_1-K_2 - K_3 + K_{q_1} - K_{q_2} - K_{q_3} + K_{p^0}\right),\\
e_2 &= \frac14 \left(+K_0 -+K_1-K_2 + K_3 - K_{q_1} + K_{q_2} - K_{q_3} + K_{p^0}\right),\\
f_2 &= \frac14 \left(-K_0 - K_1+K_2 - K_3 - K_{q_1} + K_{q_2} - K_{q_3} + K_{p^0}\right),\\
e_3 &= \frac14 \left(+K_0 + K_1+K_2 - K_3 - K_{q_1} - K_{q_2} - K_{q_3} + K_{p^0}\right),\\
f_3 &= \frac14 \left(-K_0 - K_1-K_2 + K_3 - K_{q_1} - K_{q_2} + K_{q_3} + K_{p^0}\right).
\end{align}
\end{subequations}}

We write the $\mathrm{SO}(4,4)$ group element in Borel gauge as\footnote{Note that the normalisation of $\sigma$ is changed compared to~\cite{Virmani:2012kw}.}
\begin{align}
\label{cosel}
V = e^{-U H_0} \cdot \left[\prod_{I=1,2,3} \left(e^{-\frac12 \log y^I H_I} e^{-x^I E_I}\right)\right]
\cdot e^{-\zeta^\Lambda E_{q_\Lambda}-\tilde{\zeta}_\Lambda E_{p^\Lambda}} \cdot e^{-\sigma E_0}.
\end{align}
Next, we will explain how the scalar fields appearing in this coset element are related to the physical quantities of the STU model.

\subsection{Four-dimensional metric and duality relations in $D=3$}

We parameterise the four-dimensional metric  as
\begin{align}
\label{43met}
ds_4^2 = -e^{2U} (dt+\omega_3)^2 + e^{-2U} ds_3^2.
\end{align}
The three-dimensional metric $ds_3^2$ in turn is given by (\ref{32met}).

The $D=3$ vector fields obtained by reduction from $D=4$ are defined by
\begin{align}
\label{redvec}
A^\Lambda = \zeta^\Lambda(dt+\omega_3) + A_3^\Lambda,
\end{align}
which also defines the scalars $\zeta^\Lambda$. As for any reduction of an $\mathcal{N}=2$ supergravity theory, the duality relations between vector and scalar fields in $D=3$ are
\begin{align}
\label{dualsigma}
d\sigma - \frac{1}{2}\left(\zeta^\Lambda d\tilde\zeta_\Lambda - \tilde\zeta_\Lambda d\zeta^\Lambda \right)= - e^{4U} \star d\omega_3
\end{align}
and
\begin{align}
\label{dualzetatilde}
-d\tilde\zeta_\Lambda = e^{2U} (\mathrm{Im} N)_{\Lambda\Sigma}  \star\left(dA_3^\Sigma+\zeta^\Sigma d\omega_3\right) + (\mathrm{Re} N)_{\Lambda\Sigma} d\zeta^\Sigma.
\end{align}

The matrix $N_{\Lambda\Sigma}$ is defined through the cubic prepotential $F(X) = -\frac{X^1X^2X^3}{X^0}$ via
\begin{align}
N_{\Lambda\Sigma} = \bar{F}_{\Lambda\Sigma} +2i \frac{(\mathrm{Im}F)_{\Lambda\Xi}(\mathrm{Im}F)_{\Sigma\Pi}X^\Xi X^\Pi}{ (\mathrm{Im}F)_{\Xi\Pi} X^\Xi X^\Pi},
\end{align}
where subscripts $F_\Lambda$ denote derivatives of $F$ with respect to $X^\Lambda$. In the gauge $X^0=1$ the scalar fields are (for $I=1,2,3$)
\begin{align}
z^I = \frac{X^I}{X^0}=X^I= x^I+i y^I. \label{xIyI}
\end{align}

In the present case these definitions imply (we lower the indices on $x^I$ for readability)
\begin{align}
(\mathrm{Re} N)_{\Lambda\Sigma} =
\begin{pmatrix}
 -2 x_1 x_2 x_3 & x_2 x_3 & x_1 x_3 & x_1 x_2 \\
 x_2 x_3 & 0 & -x_3 & -x_2 \\
 x_1 x_3 & -x_3 & 0 & -x_1 \\
 x_1 x_2 & -x_2 & -x_1 & 0
\end{pmatrix},
\end{align}
and
\begin{align}
(\mathrm{Im} N)_{\Lambda\Sigma} =
\begin{pmatrix}
\frac{-x_3^2 y_1^2 y_2^2-x_1^2 y_3^2 y_2^2-x_2^2 y_1^2 y_3^2-y_1^2 y_2^2 y_3^2}{y_1 y_2 y_3} & \frac{x_1 y_2 y_3}{y_1} & \frac{x_2
   y_1 y_3}{y_2} & \frac{x_3 y_1 y_2}{y_3} \\
 \frac{x_1 y_2 y_3}{y_1} & -\frac{y_2 y_3}{y_1} & 0 & 0 \\
 \frac{x_2 y_1 y_3}{y_2} & 0 & -\frac{y_1 y_3}{y_2} & 0 \\
 \frac{x_3 y_1 y_2}{y_3} & 0 & 0 & -\frac{y_1 y_2}{y_3}
\end{pmatrix},
\end{align}
with inverse
\begin{align}
((\mathrm{Im} N)^{-1})_{\Lambda\Sigma} =\frac1{y_1y_2y_3}
\begin{pmatrix}
-1 & -x_1 & -x_2 & -x_3 \\
 -x_1 & -x_1^2-y_1^2 & -x_1 x_2 & -x_1 x_3 \\
 -x_2 & -x_1 x_2 & -x_2^2-y_2^2 & -x_2 x_3 \\
 -x_3 & -x_1 x_3 & -x_2 x_3 & -x_3^2-y_3^2
\end{pmatrix}.
\end{align}

\section{Two-dimensional fields for the four-charge black hole}
\label{app:Scalars}

In this appendix we show how to obtain the four-charge solution of Cveti\v{c}--Youm from $\cV(t)$ and $V$ that were constructed in section~\ref{sec:4charge}.

The first thing to do is to change coordinates on the two-dimensional base. This is done by parameterizing the pole values of the spectral parameter through
\begin{subequations}
\bea
t_1 &=& \frac{(u-c)(1+v)}{\sqrt{(u^2-c^2)(1-v^2)}}, \\
t_2 &=& \frac{(u+c)(1+v)}{\sqrt{(u^2-c^2)(1-v^2)}}.
\eea
\end{subequations}
As a next step we change from the prolate spherical coordinates $(u,v)$ to the Boyer-Lindquist coordinates $(r,x)$ defined by
\begin{align}
u  = r-m, \qquad
v = x.
\end{align}
The constants $\zeta$ and $c$ that appear in the parameterisations of the pole and residue vectors are conveniently given in terms of $m$ and $a$ as
\begin{align}
\zeta = \frac{c-m}{a}, \qquad \qquad
c = \sqrt{m^2 - a^2}.
\end{align}
Now we introduce the abbreviations
\be
\label{Ds}
\Delta = \frac{r^2 + a^2 x^2 - 2 m r}{r^2 + a^2 x^2}, \qquad \qquad \sKerr = - \frac{2 m a x}{r^2 + a^2 x^2}.
\ee
We again stress the factor of 2 for $\s$ for Kerr compared to~\cite{Virmani:2012kw}.  Using the conformal factor \eqref{conf4charge}, the three-dimensional base metric is here found to be
\begin{align}
ds_3^2 = \frac{r^2 - 2mr + a^2 x^2}{r^2-2mr+a^2}dr^2 + (r^2 - 2mr + a^2 x^2)\frac{dx^2}{1-x^2} + (1-x^2)(r^2-2mr+a^2) d\varphi^2.
\label{3dmet}
\end{align}
We have fixed the normalization factor in \eqref{conf4charge} to be $k_\rom{BM} = - 4c^2\frac{(1+\zeta^2)^2}{(1-\zeta^2)^4}=-\frac{m^2 a^4}{c^2(m-c)^2}$ by the requirement of asymptotic flatness.

The presentation of the rest of the fields below is closely related to that of \cite{STU3}. The scalar fields $x_I$ of (\ref{xIyI}) are given by
\begin{subequations}
\bea
x_1 &=& \frac{(c_{01}  s_{23} - s_{01} c_{23})\sKerr}{h_2 h_3 + s_{23}^2 \sKerr^2}, \\
x_2 &=& \frac{(c_{02}  s_{13} - s_{02} c_{13})\sKerr}{h_1 h_3 + s_{13}^2 \sKerr^2}, \\
x_3 &=& \frac{(c_{03}  s_{12} - s_{03} c_{12})\sKerr}{h_1 h_2 + s_{12}^2 \sKerr^2}.
\eea
\end{subequations}
Introducing in addition the shorthand
\begin{subequations}
\bea
h_i &=&(c_i^2 - s_i^2\Delta) \\
c_{i_1 \ldots i_n} &=& \cosh \delta_{i_1} \ldots \cosh \delta_{i_n} \\
s_{i_1 \ldots i_n} &=& \sinh \delta_{i_1} \ldots \sinh \delta_{i_n}
\eea
\end{subequations}
the scalar fields $y_I$ of (\ref{xIyI}) are found to be
\begin{subequations}
\bea
y_1 &=& \frac{W}{h_2 h_3 + s_{23}^2 \sKerr^2} \\
y_2 &=& \frac{W}{h_1 h_3 + s_{13}^2 \sKerr^2} \\
y_3 &=& \frac{W}{h_1 h_2 + s_{12}^2 \sKerr^2},
\eea
\end{subequations}
where
\bea
\label{Weq}
W^2 &=& h_0 h_1 h_2 h_3 + \sKerr^2 \big{(}2 c_{0123}s_{0123} - (s_{012}^2 + s_{013}^2 + s_{023}^2 + s_{123}^2 + 4 s_{0123}^2) \Delta \nn  \\
 & &  + 2 s^2_{0123}\Delta^2\big{)}  + s^2_{0123} \sKerr^4.
\eea
In terms of (\ref{Weq}) and (\ref{Ds}) the dilaton of the $D=4$ to $D=3$ reduction is given by
\be
e^{2U} = \frac{\Delta}{W}.
\ee
The dual of the Kaluza--Klein vector of the reduction reads
\be
\sigma = \frac{\s_\rom{Kerr}}{2W^2}\left\{ c_{0123}\left[2 + (1-\Delta)\left(\sum_{i=0}^{3}s_i^2\right)\right] + s_{0123}\left[ \left(2 +\sum_{i=0}^{3}s_i^2\right)(\Delta^2  -\Delta + \sKerr^2)- 2 \Delta\right]\right\}.
\ee

The scalars coming from the vector multiplets are
\begin{subequations}
\bea
\tilde \zeta_0 &=& \frac{\sKerr}{W^2} \left[h_0 (s_0 c_{123} - c_0 s_{123} \Delta) + s_0 c_0 s_{0123} \sKerr^2\right] ,\\
\zeta^1 &=& \frac{\sKerr}{W^2} \left[h_1 (s_1 c_{023} - c_1 s_{023} \Delta) + s_1 c_1 s_{0123} \sKerr^2\right], \\
\zeta^2 &=& \frac{\sKerr}{W^2} \left[h_2 (s_2 c_{013} - c_2 s_{013} \Delta) + s_2 c_2 s_{0123} \sKerr^2\right], \\
\zeta^3 &=& \frac{\sKerr}{W^2} \left[h_3 (s_3 c_{012} - c_3 s_{012} \Delta) + s_3 c_3 s_{0123} \sKerr^2\right],
\eea
\end{subequations}
and
\begin{subequations}
\bea
\zeta^0 &=& + \left\{\frac{c_0}{s_0} - \frac{1}{s_0 W^2} (c_0 h_1 h_2 h_3 + (s_0 c_{123} - c_0 s_{123} \Delta)s_{123} \sKerr^2) \right\}, \\
\tilde \zeta_1 &=& - \left\{\frac{c_1}{s_1} - \frac{1}{s_1 W^2} (c_1 h_0 h_2 h_3 + (s_1 c_{023} - c_1 s_{023} \Delta)s_{023} \sKerr^2) \right\}, \\
\tilde \zeta_2 &=& - \left\{\frac{c_2}{s_2} - \frac{1}{s_2 W^2} (c_2 h_0 h_1 h_3 + (s_2 c_{013} - c_2 s_{013} \Delta)s_{013} \sKerr^2) \right\}, \\
\tilde \zeta_3 &=& - \left\{\frac{c_3}{s_3} - \frac{1}{s_3 W^2} (c_3 h_0 h_1 h_2 + (s_3 c_{012} - c_3 s_{012} \Delta)s_{012} \sKerr^2) \right\}.
\eea
\end{subequations}

Upon substituting the expressions for $\sKerr$ and $\Delta$ and after performing the dualizations using (\ref{dualsigma}) and (\ref{dualzetatilde}), the above expressions take the following form
\begin{subequations}
\bea
x_1 &=& 2m a x\frac{s_{01} c_{23} - c_{01} s_{23}}{r_2 r_3 + a^2 x^2}, \\
x_2 &=& 2m a x\frac{s_{02} c_{13} - c_{02} s_{13}}{r_1 r_3 + a^2 x^2} ,\\
x_3 &=& 2m a x\frac{s_{03} c_{12} - c_{03} s_{12}}{r_1 r_2 + a^2 x^2},
\eea
\end{subequations}
where $r_i = r + 2 m s_i^2$, and
\begin{subequations}
\bea
y_1 &=& \frac{\tilde W}{r_2 r_3 + a^2 x^2} ,\\
y_2 &=& \frac{\tilde W}{r_1 r_3 + a^2 x^2} ,\\
y_3 &=& \frac{\tilde W}{r_1 r_2 + a^2 x^2}.
\eea
\end{subequations}
with $\tilde{W}^2 := (r^2 + a^2 x^2)^2 W^2$ given below in (\ref{Wtilde}).
The scalars appearing in (\ref{redvec}) are
\begin{subequations}
\bea
\zeta^0 &=&  \frac{2 m c_0 s_0 (r_1 r_2 r_3 + r a^2 x^2) + 4 a^2 m^2 x^2 e_0}{\tilde W^2}, \\
\zeta^1 &=& - 2 m a x \frac{(s_1 c_{023} - c_1 s_{023})(r r_1 + a^2 x^2)+ 2 m c_1 s_{023} r_1}{\tilde W^2}, \\
\zeta^2 &=& - 2 m a x \frac{(s_2 c_{013} - c_2 s_{013})(r r_2 + a^2 x^2)+ 2 m c_2 s_{013} r_2}{\tilde W^2}, \\
\zeta^3 &=& - 2 m a x \frac{(s_3 c_{012} - c_3 s_{012})(r r_3 + a^2 x^2)+ 2 m c_3 s_{012} r_3}{\tilde W^2},
\eea
\end{subequations}
where
\be
e_0 = (c_0^2 + s_0^2)c_{123} s_{123} - c_0 s_0 (s_{12}^2  + s_{23}^2 + s_{13}^2 + 2s_{123}^2).
\ee
The three dimensional one-forms read with (\ref{dualsigma}) and (\ref{dualzetatilde})
\be
\omega_3  = 2 a m (1-x^2) \frac{(c_{0123} r  - (r -2m)s_{0123})}{r^2 - 2m r + a^2 x^2} d\varphi,
\ee
and
\bea
A_3^0 &=& - 2 a m (1-x^2) \frac{(s_0 c_{123} r  - (r -2m) c_0 s_{123})}{r^2 - 2m r + a^2 x^2} d\varphi, \\
A_3^1  &=&  2  m s_1 c_1 x \frac{r^2 + a^2 - 2m r}{r^2 - 2m r + a^2 x^2} d\varphi, \\
A_3^2  &=&  2  m s_2 c_2 x \frac{r^2 + a^2 - 2m r}{r^2 - 2m r + a^2 x^2} d\varphi, \\
A_3^3  &=&  2  m s_3 c_3 x \frac{r^2 + a^2 - 2m r}{r^2 - 2m r + a^2 x^2} d\varphi.
\eea
Finally,
\bea
\label{Wtilde}
\tilde W^2 &=&
r_0 r_1 r_2 r_3 + a^4 x^4 + a^2 x^2 [2r^2 + 2 m r (s_0^2 + s_1^2 + s_2^2 + s_3^2) \nn \\
 & & + 8 m^2 c_{0123} s_{0123} - 4m^2 (s^2_{012} + s^2_{123} + s^2_{023} s^2_{013} + 2 s^2_{0123})].
\eea
Using these expressions the four-dimensional metric and the various matter fields can be readily obtained by substitution into (\ref{43met}) and (\ref{redvec}).
In these expressions $a$ is the bare rotation parameter and $m$ is the bare mass parameter.

\end{document}